\documentclass[reprint,amsmath,amssymb,aps]{revtex4-2}
\usepackage{graphicx}
\usepackage{subfig}
\usepackage{dcolumn}
\usepackage{bm}
\usepackage{xcolor}
\DeclareMathOperator{\sech}{sech}

\begin{document}

\title{Quasi-normal $f$-modes of anisotropic quark stars in full general relativity}

\author{Sushovan Mondal}
\email{smondal@imsc.res.in}
\affiliation{The Institute of Mathematical Sciences, HBNI, Taramani, Chennai 600113, India.}
\affiliation{Homi Bhabha National Institute, Training School Complex, Anushakti Nagar, Mumbai 400094, India.}

\author{Manjari Bagchi}
 \email{manjari@imsc.res.in}
\affiliation{The Institute of Mathematical Sciences, HBNI, Taramani, Chennai 600113, India.}
\affiliation{Homi Bhabha National Institute, Training School Complex, Anushakti Nagar, Mumbai 400094, India.}

\date{\today}

\defcitealias{mondal2024prd}{Paper-I}

\begin{abstract}
We investigate the fundamental mode, or $f$-mode, oscillations of anisotropic quark stars within the framework of full general relativity. We consider two different equations of state (EOSs), one is the MIT bag model EOS for non-interacting quark matter, and the other is EOS-A, which incorporates inter-quark interactions. Our study systematically examines the impact of pressure anisotropy on the equilibrium structure, as well as on the frequencies and damping times of $f$-mode oscillations in quark stars. Our results confirm that the $f$-mode frequency scales linearly with the square root of the average density, with anisotropy influencing both the slope and intercept of this relation. This behavior is consistent with our previous findings for neutron stars. The dependence of the $f$-mode frequency on total mass reveals distinct trends based on the relative dominance of tangential and radial pressure. When the tangential pressure exceeds the radial pressure, the frequency increases monotonically with mass, exhibiting rapid growth for massive quark stars. Conversely, when the radial pressure dominates, the frequency also increases with mass; however, in cases where the radial pressure is significantly greater than the tangential pressure, the frequency decreases as mass increases. Additionally, we find that for low-mass quark stars, stronger tangential pressure leads to an increase in frequency, while beyond a threshold mass, a further increase in tangential pressure results in a decrease in frequency. For the chosen range of anisotropic strengths, the frequency varies between 1.3 kHz and 2.3 kHz for the MIT bag EOS and between 1.8 kHz and 3.4 kHz for EOS-A. Furthermore, we find that the normalized damping time follows a linear trend with compactness, with anisotropic strength affecting both the slope and intercept of this relation. For a fixed stellar mass, an increase in tangential pressure relative to radial pressure reduces the damping time, whereas a decrease in tangential pressure significantly increases it. The damping time ranges from 83 ms to 900 ms for the MIT bag EOS and from 60 ms to 761 ms for EOS-A. Finally, we present semi-empirical expressions for both the frequency and damping time as functions of mass, radius, and anisotropic strength. The frequency exhibits a cubic polynomial dependence on anisotropy, while the damping time follows a quartic dependence. Our results provide new insights into the role of anisotropy in quark star oscillations and may aid in constraining the equation of state of dense matter through future gravitational wave observations.

\end{abstract}

\maketitle

\section{\label{intro} Introduction}
The concept of neutron stars was first introduced by \citet{baade1934pnasa, baade1934pnasb} in the early 1930s. However, the first direct detection of a neutron star occurred in 1967, when Jocelyn Bell first identified a pulsar \cite{hewish1968nature}. Pulsars are rapidly rotating neutron stars emitting electromagnetic beams along its magnetic axis that falls onto the earth once or twice in its one rotation.

Neutron stars, with masses typically ranging from $1-2~{\rm M_\odot}$ and diameters of about 20 km, are among the densest known astrophysical objects and belong to the class of `compact stars'. These extreme conditions provide unique natural laboratories for investigating matter at densities far exceeding those achievable in terrestrial experiments. Understanding the physical properties and behavior of such dense matter remains one of the most compelling challenges in both astrophysics and particle physics.

A wide variety of equations of state (EOSs) have been proposed to describe the matter within these compact stars \cite{miller2020assl, akmal1998PRC, douchin2001AA, lattimer1991NPA, lattimer2021ARNPS}. Based on the constituent particles, these can be baryonic stars containing different types of baryons or hybrid stars having free quarks near the cores. All of such stars are commonly referred to as neutron stars.

\citet{itoh1970ptp} first suggested that compact stars composed entirely of quark matter could exist in hydrostatic equilibrium. It has further been argued that quark matter might represent the true ground state of matter, as it is potentially more stable than ${}^{56}\text{Fe}$ nuclei \cite{bodmer1971prd, witten1984prd}. This hypothesis strengthens the idea of the existence of compact stars composed entirely of quark matter, conventionally referred to as quark stars.

Although quark stars remain hypothetical due to the lack of direct observational evidence, several objects have been proposed as potential quark star candidates. A recent study suggests that the low-mass central object of the supernova remnant HESS J1731$-$347 could be a quark star \cite{doroshenko2022NA, horvath2023light}, as such a low-mass neutron star is unlikely to form through conventional core-collapse supernova mechanisms. Additionally, other candidates, such as PSR B$0943+10$ \cite{Yue2006APJ} and PSR J$0205+6449$ \cite{weber2005ppnp}, have also been speculated to be quark stars.

It is more common to assume the pressure inside neutron stars to be isotropic. However, there exist various physical processes that can induce pressure anisotropy. Some of the proposed processes are superfluidity \cite{ruder1972ARAA, Hoffberg1970PRL, Sokolov1980SJETP}, pion condensation \cite{sawyer1972prl}, skyrmionic interactions \cite{nelmes2012PRD}, etc. Although these processes are mostly relevant for neutron stars, there are anisotropy causing processes that can be relevant for quark stars too, one example being the presence of viscosity \cite{barreto1992equation, barreto1993exploding}. \citet{Yazadjiev2012PRD} modeled magnetars in general relativity using a nonperturbative approach, considering anisotropic fluid behavior due to magnetic fields. Additionally, elasticity in compact stars can be described in terms of local anisotropy \cite{alho2022compact, Karlovini2002CQG}. Further studies on anisotropic pressure in self-gravitating objects are discussed in \citet{Herrera1997prpt} and references therein.

In addition to conventional observational studies of compact stars in the electromagnetic wavelengths, presently there is increased interest to observe these sources in the gravitational waves. Observations of gravitational waves from compact binary mergers by LIGO \cite{abbot2017PRL, abbot2020ApJL} have provided a novel way to probe the interior structure of neutron stars and other compact objects. Non-radial oscillation of compact stars are one of the prime candidate of gravitational waves. Phenomena such as pulsar glitches \cite{wilson2024prd}, non-spherical supernova explosions \cite{kumar2024mnras}, and coalescence of compact stars \cite{kuan2022prd,stergioulas2011,bauswein2015,kabir2020,lioutas2021} are among of the proposed causes of such non-radial, especially $f$-mode oscillations. These are the sources of gravitational wave in burst-like emission. Other than these sources, f-mode oscillation can also generate continuous gravitational waves by spinning compact stars. The instability of the f-mode  sets in only above a critical spin rate, via the well-known Chandrasekhar-Friedman-Schutz (CFS) \cite{friedman1975,chandra1970} mechanism. In a rapidly rotating (millisecond-period) newborn neutron star, the CFS instability can drive the f-mode to grow and emit continuous gravitational waves \cite{surace2015,passamonti2012,dong2024MNRAS}.

The study of non-radial oscillations were done by various researchers for over the last fifty years, sometimes under Cowling approximation and sometimes with within full general relativistic framework \cite{thorne1967apj,lindblom1983apj, detweiler1985apj, comer1999PRD, sotani2002prd, minutti2003MNRAS, mondal2024prd}. Cowling approximation assumes that metric perturbations are zero. While this simplifies the calculations, the omission of metric perturbations may introduce non-negligible errors \cite{sotani2020prd}. Additionally, since the Cowling approximation neglects metric perturbations, it fails to account for energy loss due to gravitational waves, making it impossible to compute damping times within this framework.     

Quasinormal modes of quark stars were first studied by \citet{yip1999apj}, who demonstrated that the $f$-modes of quark stars differ from those of neutron stars and that quark stars are more efficient gravitational wave radiators than pure neutron stars. \citet{kojima2002ptp} showed that the $f$-mode frequency and damping time can be used to distinguish between quark stars and neutron stars. \citet{sotani2003prd} derived an empirical formula relating the $f$-mode frequency of gravitational waves to the parameters of the equation of state for quark stars. Similarly, \citet{benhar2007grg} argued that knowledge of the $f$-mode frequency, along with the mass and radius of the source, can help differentiate between neutron stars and quark stars.

These studies shed light on the importance of $f$-modes for understanding the properties of quark stars and distinguishing them from neutron stars. However, all of the aforementioned investigations assumed isotropic pressure inside the quark star. Some studies of quasinormal modes in anisotropic quark stars have been conducted \cite{zhang2024fass, pretel2024JCAP}, but they employed the Cowling approximation. 

Recently, we investigated the $f$-mode oscillations of anisotropic, non-rotating neutron stars within the framework of full general relativity \cite{mondal2024prd}, hereafter referred to as Paper-I. Our study revealed that the $f$-mode frequency exhibits a linear relationship with the square root of the average density of the neutron star, with the slope of the linear fit depending on the strength and nature of the anisotropy. When the tangential pressure exceeds the radial pressure (positive anisotropy), the $f$-mode frequency increases, while a negative anisotropy leads to a reduction in frequency. Additionally, we observed that for lower mass neutron stars, the frequency increases linearly with mass, whereas for higher mass stars, the frequency rises nonlinearly with mass. This nonlinear behavior is more prominent in configurations where the tangential pressure is larger than the radial pressure. Our findings highlighted that anisotropy significantly impacts the oscillatory properties of neutron stars, including their damping times. For mildly anisotropic neutron stars near the maximum stable mass, a slight increase in damping time was observed, although damping times generally decrease with increasing mass. Furthermore, we noted that softer EOSs lead to higher $f$-mode frequencies and shorter damping times for a fixed anisotropic strength. These results show the critical role anisotropy plays in shaping the oscillatory properties of compact stars. Extending these studies to quark stars could provide deeper insights into the interplay between anisotropy and the dense matter EOS.

Therefore, in the present, we extend the analysis of anisotropic compact stars to quark stars within full general relativity. We focus on the $ l = 2 $ (quadrupole) $f$-mode, which is the dominant non-radial oscillation that couples most efficiently to gravitational radiation and is the primary target in gravitational-wave asteroseismology. We numerically compute the $f$-mode frequency and the associated damping time for anisotropic quark stars, considering two different equations of state (EOSs), and analyze the impact of anisotropy on these oscillation properties. Additionally, we derive approximate analytical expressions for both the frequency and damping time as functions of the stellar mass, radius, and anisotropic strength parameter. These analytical relations might be useful tools for future studies involving gravitational wave signatures from compact objects.

This paper is organized as follows. In Sec.~\ref{sec:equilib}, we discuss the equilibrium structure of anisotropic quark stars. Sec.~\ref{result:fmode} explores the impact of anisotropy on $f$-mode frequencies and damping times for different quark star configurations. In Sec.~\ref{sec:analytic}, we present approximate analytical expressions for the $f$-mode frequency and damping time as functions of mass, radius, and anisotropy strength. Finally, in Sec.~\ref{conclusion}, we summarize our findings and outline potential directions for future research.

\section{Equilibrium Anisotropic Configuration of Quark Stars in General Relativity}\label{sec:equilib}

The equilibrium equations governing the structure of stars having anisotropic pressure within are described by the modified Tolman-Oppenheimer-Volkoff (TOV) equations, as thoroughly outlined in \citetalias{mondal2024prd}. In the present work, we follow the same conventions and notations as in \citetalias{mondal2024prd}, unless otherwise specified. As our primary focus is on the $f$-mode oscillations of quark stars, we next describe the EOSs for the strange quark matter employed to analyze the oscillatory behavior of these stars. Additionally, we briefly review the anisotropy parameter, which is identical to that introduced in \citetalias{mondal2024prd}.

\subsection{Equation of State of Quark Matter}\label{sec:eos}

\subsubsection{Noninteracting Quark Matter}\label{nonintqm}
The simplest EOS for quark stars is provided by the MIT bag model. In this framework, all three flavors of quarks (up, down, and strange quarks) are treated as non-interacting particles confined within a hypothetical bag. The radial pressure in this model is expressed as:
\begin{equation}\label{MITpr}
	p_r = -B + \sum_{i = u,d,s} p^i,
\end{equation}
where $p^i$ denotes the pressure contribution from each quark flavor, counterbalanced by the total external bag pressure $B$. 

The total energy density $\rho$ of the deconfined quarks inside the bag is given by:
\begin{equation}\label{MITrho}
	\rho = \sum_{i = u,d,s} \rho^i + B,
\end{equation}
where $\rho^i = 3 p^i$ represents the energy density of each quark flavor. Substituting this into the pressure equation, the EOS for quark matter can be simplified and written as:
\begin{equation}\label{MITprfin}
	p_r = \frac{1}{3} (\rho - 4 B).
\end{equation}

In this study, the bag constant is fixed at $B = 56 ~ \text{MeV} \, \text{fm}^{-3}$, ensuring consistency with prior works \cite{zdunik2002AA, zudnil2001AA, sotani2003prd}. 

\subsubsection{Interacting Quark Matter}\label{intqm}

A family of alternative EOSs for strange quark matter is given by Dey et al. \cite{dey1998plb}, later improved by Bagchi et al. \cite{bagchi2006aa}. These EOSs are derived using a relativistic Hartree-Fock calculation for strange quark matter, incorporating a modified form of the phenomenological inter-quark interaction known as the Richardson potential \cite{richardson1979plb}. The modified Richardson potential is given by:
\begin{align}\label{richpot}
	V_{ij}  = &\frac{12 \pi}{7}\left[\frac{1}{\ln\left(1 + \frac{(\mathbf{k_i} - \mathbf{k_j})^2}{\Lambda}\right)} - \frac{\Lambda^2}{(\mathbf{k_i} - \mathbf{k_j})^2} + \frac{{\Lambda^{\prime}}^2}{(\mathbf{k_i} - \mathbf{k_j})^2}\right] \nonumber\\
	& \times \frac{1}{(\mathbf{k_i} - \mathbf{k_j})^2},
\end{align}  
where $\Lambda$ and $\Lambda^{\prime}$ characterize the asymptotic freedom and confinement properties, respectively, and $(\mathbf{k_i} - \mathbf{k_j})$ represents the momentum transfer between the $i$-th and the $j$-th quarks. Applying the same potential in the baryonic sector \cite{bagchibaryon2004, bagchibaryon2006}, it was found that the realistic values of $\Lambda ^{\prime}$ can be taken in the range of 300-350 MeV, while the value of $\Lambda$ is better fixed at 100 MeV.

In a medium, this bare potential undergoes screening due to pair creation and infrared divergence. As a result, $(\mathbf{k_i} - \mathbf{k_j})^2$ in Eq.~(\ref{richpot}) is replaced by $[(\mathbf{k_i} - \mathbf{k_j})^2 + D^{-2}]$, where $D$ denotes the screening length. The inverse screening length $(D^{-1})$ is expressed at zero temperature to the lowest order as:
\begin{equation}\label{screenlen}
	(D^{-1})^2 = \frac{2 \alpha_0}{\pi} \sum_{i = u,d,s} k_i^f\sqrt{\left(k_i^f\right)^2 +  \mathcal{M}_{{\rm cur}, i}^2} \, ,
\end{equation}
where $k_i^f$ is the Fermi momentum of the $i$ quark,  $\mathcal{M}_{{\rm cur}, i}$ is the current mass of the $i$ quark, and $\alpha_0$ is the perturbative quark-gluon coupling. 

This model accounts for chiral symmetry restoration at high densities by considering the density-dependent quark mass $M_i$, which decreases with increasing density as follows:
\begin{equation}\label{QM}
	\mathcal{M}_i = \mathcal{M}_{{\rm cur}, i}+ \mathcal{M}_Q \sech \left(\frac{n_B}{N n_0}\right), \quad \text{for } i = u,d,s,
\end{equation}
where $n_B = (n_u + n_d + n_s)/3$ is the baryon number density, $n_0 = 0.17~\rm fm^{-3}$ is the nuclear matter saturation density, and $n_u$, $n_d$, and $n_s$ denote the number densities of up, down, and strange quarks, respectively. The parameters $\mathcal{M}_Q$ and $N$ are tuned to ensure that the minimum energy per baryon for strange quark matter remains below that of the most stable element, ${}^{56}\rm Fe$ (930 MeV). This minimum energy per baryon is calculated at the surface of the star, where the radial pressure vanishes. The current quark masses are taken as $\mathcal{M}_{{\rm cur}, u} = 4~\rm MeV$, $\mathcal{M}_{{\rm cur}, d} = 7~\rm MeV$, and $\mathcal{M}_{{\rm cur}, s} = 150~\rm MeV$.

To maintain the charge neutrality of the matter, some electrons also needed to be present in the matter. While the Fermi momenta ($k_i^f$) of the $i$th quark is related to its number density $(n_i)$ as $k_i^f = (n_i \pi^2)^{1/3}$, the relation between the Fermi momentum and number density of the electron is $k_e^f = (3 n_e \pi^2)^{1/3}$. The values of Fermi momenta and chemical potentials of various particles are obtained in such a way that the charge neutrality and beta equilibrium conditions are satisfied throughout the matter. Then, using the laws of thermodynamics, the radial pressure $p_r$ for various values of the density $\rho$, i.e., the EOS, is obtained. Different values of the model parameters $\Lambda$, $\alpha_0$, $\mathcal{M}_Q$, and $N$ give different EOSs. Out of various EOSs given by \citet{bagchi2006aa}, we use EOS-A in the present work. EOS-A corresponds to $\Lambda' = 350 ~ \rm MeV$, $N = 3.0$, $\alpha_0 = 0.55$, and $\mathcal{M}_Q = 325 ~ \rm MeV$. 
 
\subsection{Ansatz for Anisotropy}

One can model the pressure anisotropy inside a star in two different ways. First, it might be possible to define two separate EOSs, one relating the radial pressure ($p_r$) with the density ($\rho$) and the other relating the tangential pressure ($p_t$) with $\rho$. The existing literature lacks the knowledge to proceed in this path. The second method is to specify the EOS for the radial direction while introducing an ansatz for the anisotropy parameter $(\chi)$ to account for the tangential pressure as $p_t = p_r + \chi$. Various popular ans\"atze for modeling $\chi$ are available in the literature \cite{bowers1974apj, horvat2011cqg}. Following \citetalias{mondal2024prd}, we adopt the ansatz proposed by \citet{horvat2011cqg}, which is expressed as:
\begin{equation} \label{quasiparam}
	\chi = \tau p_r \mu ,
	\end{equation} which we call as the `Horvat ansatz'. In Eq. (\ref{quasiparam}) $\mu$ is the dimensionless local compactness of the star defined as $\mu = 2 m / r$, where $m$ is the mass enclosed within the spherical region of radius $r$ inside the star. $\tau$ is a dimensionless free parameter that controls the degree of the anisotropy.

The form of the anisotropy parameter in Eq.~(\ref{quasiparam}) exhibits two key advantages. Firstly, the anisotropy parameter vanishes at the center of the star because the compactness scales as $\mu \sim r^2$ when $r \to 0$, ensuring the regularity of $\chi$. Secondly, this formulation aligns well with astrophysical expectations that, in the non-relativistic regime where $\mu \ll 1$, the impact of pressure anisotropy is negligible.

\subsection{Equilibrium quark star models}

Most of the earlier studies of anisotropic neutron stars with the Horvat ansatz  chose the values of $\tau$ within the range $-2 \leq \tau \leq 2$ \cite{mondal2024prd, doneva2012prd, folomeev2018prd, silva2015cqg}. As a starting point, we use the same range for the values of $\tau$ for anisotropic quarks stars too. We then solve the modified TOV equations numerically as described in \citetalias{mondal2024prd} to obtain total mass ($M$) and corresponding radius ($R$) for different chosen values of the central densities ($\rho_c$).

\begin{figure*}
	\centering
	\includegraphics[width=0.48\textwidth]{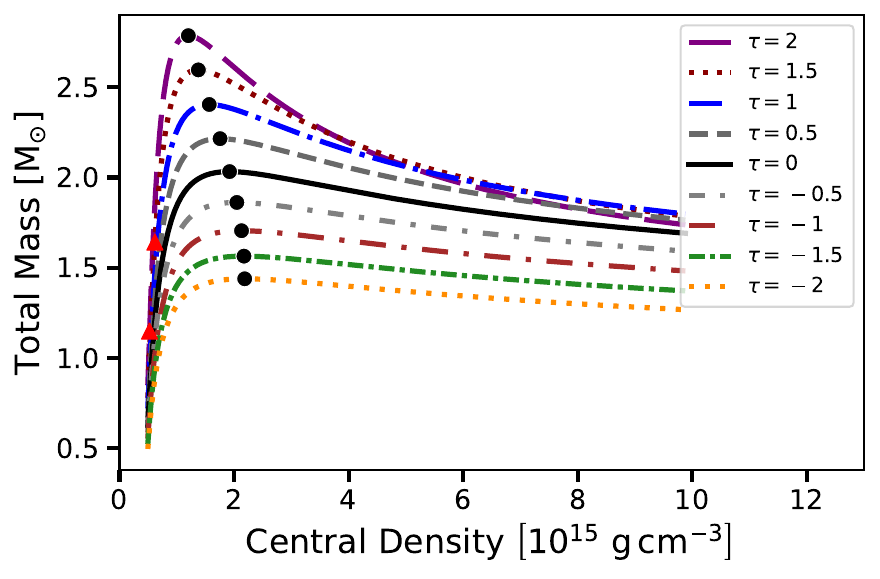}
	\hfill
	\includegraphics[width=0.48\textwidth]{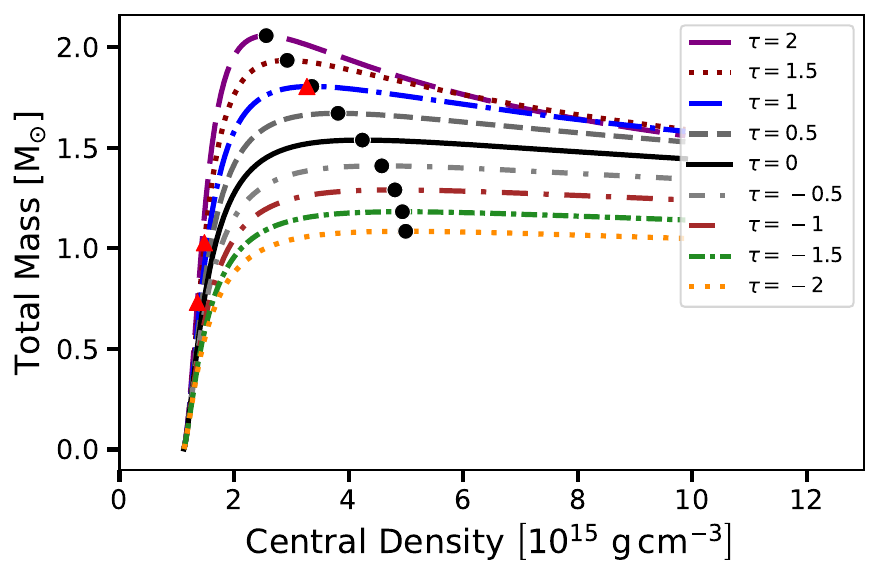}
	\caption{Mass ($M$)-Central density ($\rho_c$) profiles for anisotropic quark stars with $-2 \leq \tau \leq 2$ for the MIT bag EOS (left panel) and EOS-A (right panel). Filled circles on each profile represent the points where $\partial M/\partial \rho_c = 0$, and the red solid triangles indicate the values of $\rho_c$ up to which $v_{st} \geq 0$ throughout the star.}
	\label{M-Rhoc}
\end{figure*}

\begin{figure*}
	\centering
	\includegraphics[width=0.48\textwidth]{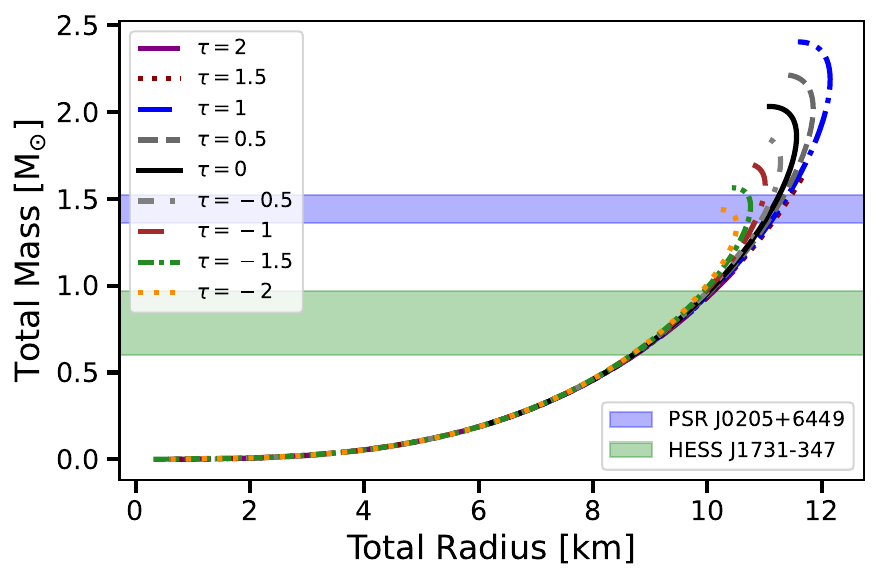}
	\hfill
	\includegraphics[width=0.48\textwidth]{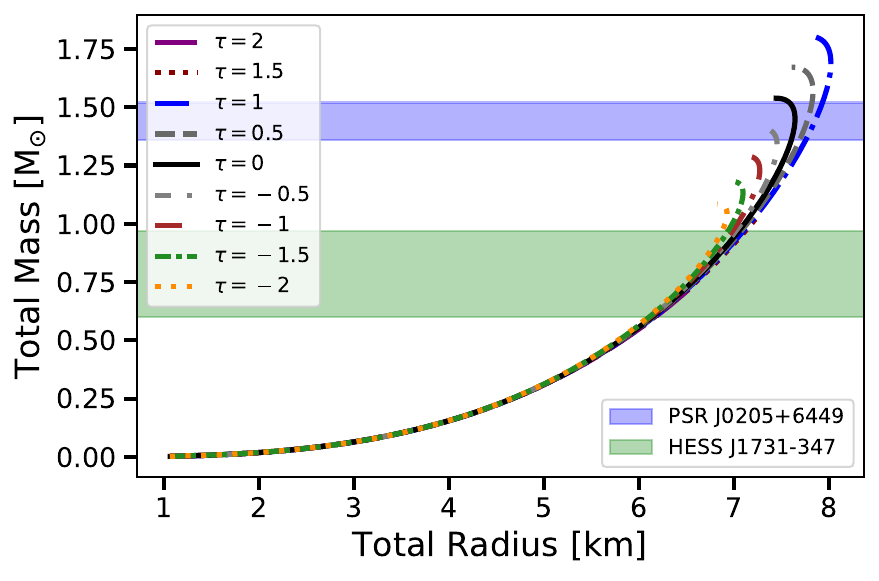}
	\caption{Mass-Radius profiles for stable anisotropic quark stars with $ -2 \leq \tau \leq 2 $, using the MIT bag EOS (left panel) and EOS-A (right panel). The horizontal shaded bands indicate the estimated mass ranges of two compact star candidates: the blue band corresponds to PSR J0205+6449 $(1.36 - 1.52~{\rm M_\odot})$ \cite{yakovlev2002}, and the green band corresponds to the central object in the supernova remnant HESS J1731$-$347 $(0.60 - 0.97~{\rm M_\odot})$ \cite{doroshenko2022NA}.}
	\label{M-R}
\end{figure*}

Fig. \ref{M-Rhoc} illustrates the total mass versus the central density profiles for anisotropic quark stars using the MIT bag EOS (left panel) and EOS-A (right panel) for several values of $\tau$. The filled circles on each profile denote the points where $\partial M / \partial \rho_c = 0$, beyond which quark stars become unstable to radial perturbations. The red solid triangles mark the maximum values of $\rho_c$ for which the sound speed in the tangential direction ($v_{st} = \sqrt{\partial p_t / \partial \rho}$) remains positive throughout the star. The positivity of the radial pressure is pre-existing in the EOSs. Similarly, Fig.~\ref{M-R} presents the mass-radius profiles for stable anisotropic quark stars with using the MIT bag EOS (left panel) and EOS-A (right panel) with the ranges of $\tau$ same as those chosen in Fig. \ref{M-Rhoc}.

For both EOSs, we observe that as the tangential pressure increases relative to the radial pressure, the maximum mass of the quark star also increases. However, for $\tau > 1$, the maximum stable mass decreases as the square of the tangential sound speed becomes negative, leading to unphysical situations. Specifically, for the MIT bag EOS, the maximum mass of an anisotropic quark star is approximately $2.45~ {\rm M_\odot}$ at $\tau = 1$. For EOS-A, the maximum stable mass is around $1.75~{\rm M_\odot}$, which also occurs at $\tau = 1$.

\begin{figure*}
	\centering
	\includegraphics[width=0.48\textwidth]{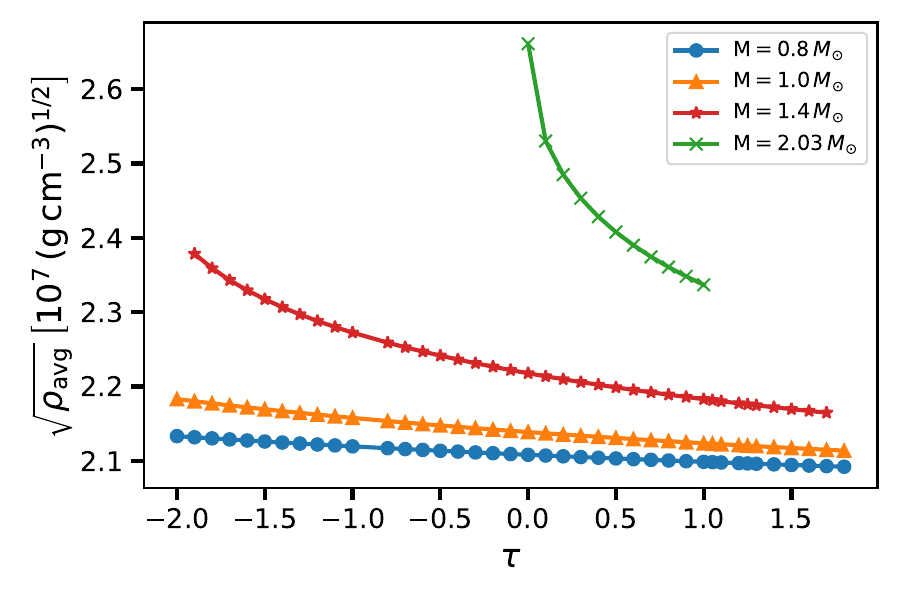}
	\hfill
	\includegraphics[width=0.48\textwidth]{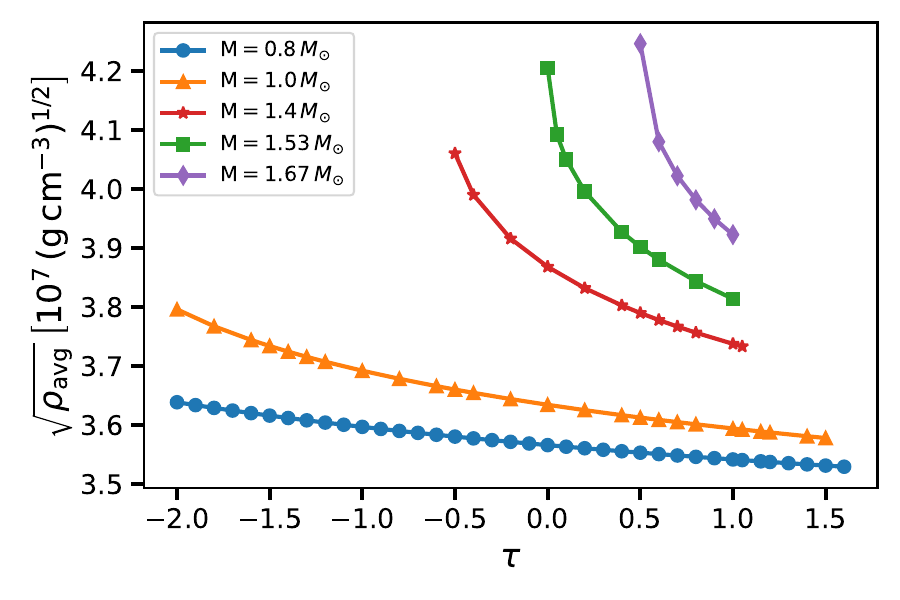}
	\caption{The variation of the square root of the average density $\sqrt{\rho_{\rm avg}}$ with the anisotropic strength $\tau$ for quark stars with different masses. The left panel corresponds to the MIT bag EOS, while the right panel represents EOS-A.}
	\label{avden-tau}
\end{figure*}

\section{Results for $f$-mode oscillations of anisotropic quark stars} \label{result:fmode}

In this section, we present the results of our investigation into the $f$-mode oscillations of anisotropic quark stars within the framework of full general relativity for both MIT bag EOS and EOS-A as described in Sec. \ref{sec:eos}. We use the analytical expressions presented in \citetalias{mondal2024prd}, as well as exactly same numerical techniques. We also use the same definitions of the frequency ($\mathcal{F}$) and damping time ($\mathfrak{T}$) of the f-modes as $\mathcal{F} = \rm{Re}(\omega) (2 \pi)^{-1} $ and $\mathfrak{T} = {\rm{Im}(\omega)}^{-1} $, respectively where $\omega$ is the complex frequency appearing in oscillation equations, whose real and imaginary parts are denoted as $\rm{Re}(\omega)$ and $\rm{Im}(\omega)$, respectively.

\subsection{Effect of anisotropy on $f$-mode frequency}

In this sub-section, first, we examine the impact of the chosen value of $\tau$ on the square root of the average density of quark stars. The average density is defined as  $\rho_{\rm avg} = M \left[ (4/3) \pi R^3 \right]^{-1}$. Fig.~\ref{avden-tau} displays the variation of $\sqrt{\rho_{\rm avg}}$ with $\tau$ for quark stars modeled using the MIT bag EOS (left panel) and EOS-A (right panel). From these plots, we see a clear trend that $\sqrt{\rho_{\rm avg}}$ decreases with the increase of $\tau$. Furthermore, the rate of decrease is more pronounced for higher-mass quark star. Quantitatively, for the MIT bag EOS, we find that $\sqrt{\rho_{\rm avg}}$ decreases by $0.4\%$, $0.7\%$, $1.56\%$, and $12.19\%$ for quark stars with masses of $0.8~{\rm M_\odot}$, $1~{\rm M_\odot}$, $1.4~{\rm M_\odot}$, and $2.03~ {\rm M_\odot}$, respectively, as $\tau$ increases from $0$ to $1$. Here, $2.03~ {\rm M_\odot}$ is the maximum mass of a stable isotropic quark star with MIT bag EOS. Similarly, for EOS-A, we observe that $\sqrt{\rho_{\rm avg}}$ decreases by $0.68\%$, $1.1\%$, $3.37\%$, and $9.30\%$ for quark stars with masses of $0.8~{\rm M_\odot}$, $1~{\rm M_\odot}$, $1.4~{\rm M_\odot}$, and $1.53 ~{\rm M_\odot}$, respectively, as $\tau$ increases from $0$ to $1$. The value $1.53~{\rm M_\odot}$ corresponds to the maximum mass of a stable isotropic $(\tau = 0)$ quark star modeled using EOS-A. It is worth noting that similar trends in the dependence of $\sqrt{\rho_{\rm avg}}$ on $\tau$ have also been reported in the context of neutron stars, as discussed in \citetalias{mondal2024prd}.

\begin{figure*}
	\centering
	\includegraphics[width=0.48\textwidth]{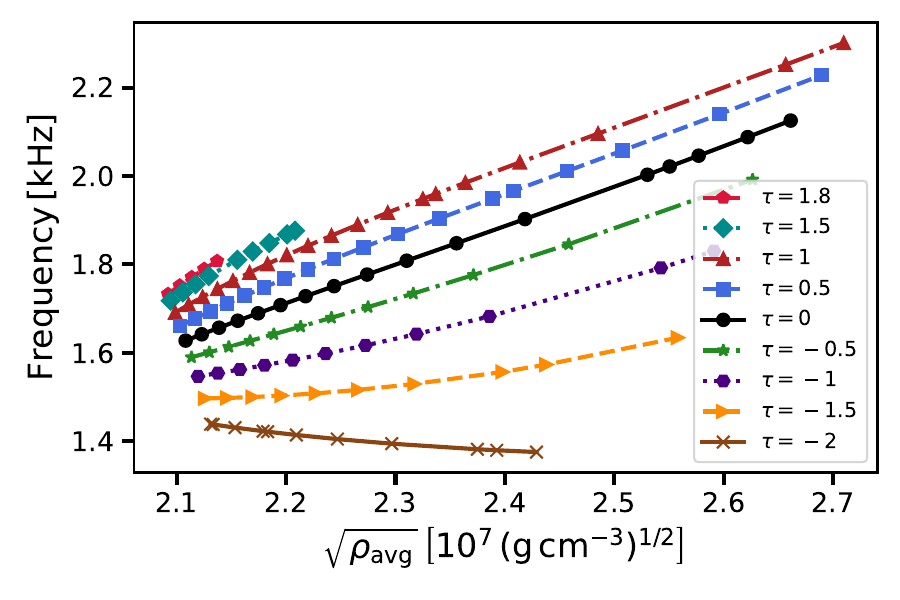}
	\hfill
	\includegraphics[width=0.48\textwidth]{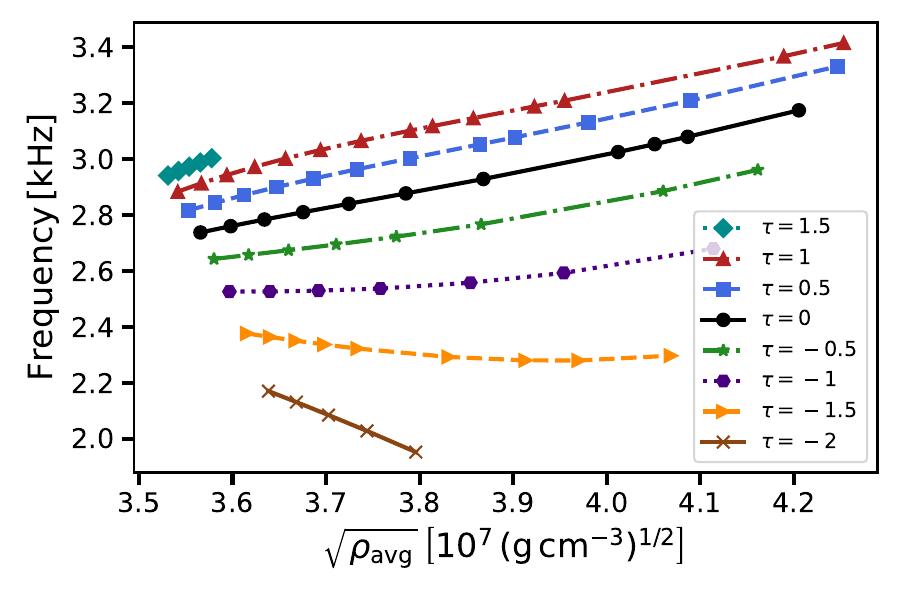}
	\caption{The variation of the $f$-mode frequency with the square root of the average density of quark stars for various values of the anisotropic strength for the MIT bag EOS (left panel) and EOS-A (right panel).}
	\label{freq_avden}
\end{figure*}

\begin{figure*}
	\centering
	\includegraphics[width=0.48\textwidth]{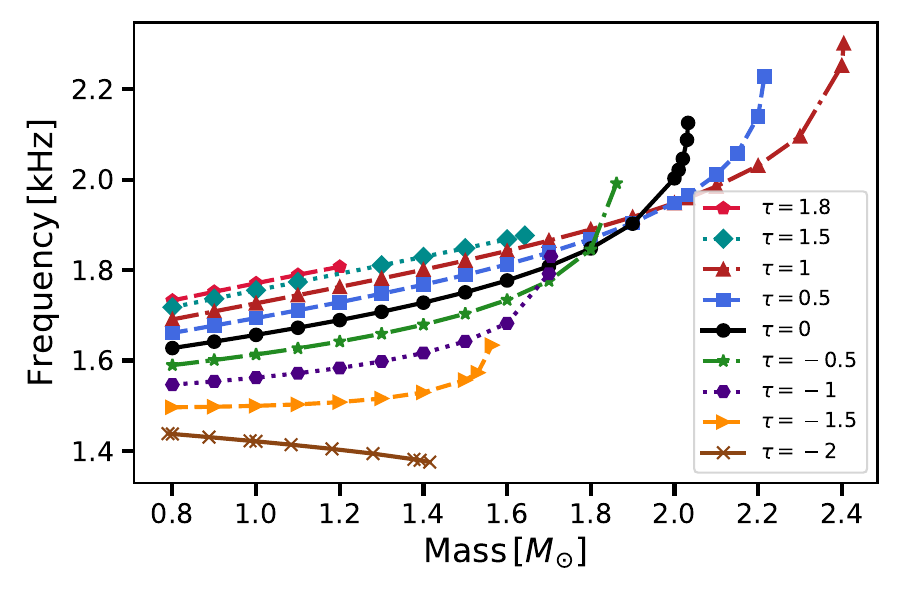}
	\hfill
	\includegraphics[width=0.48\textwidth]{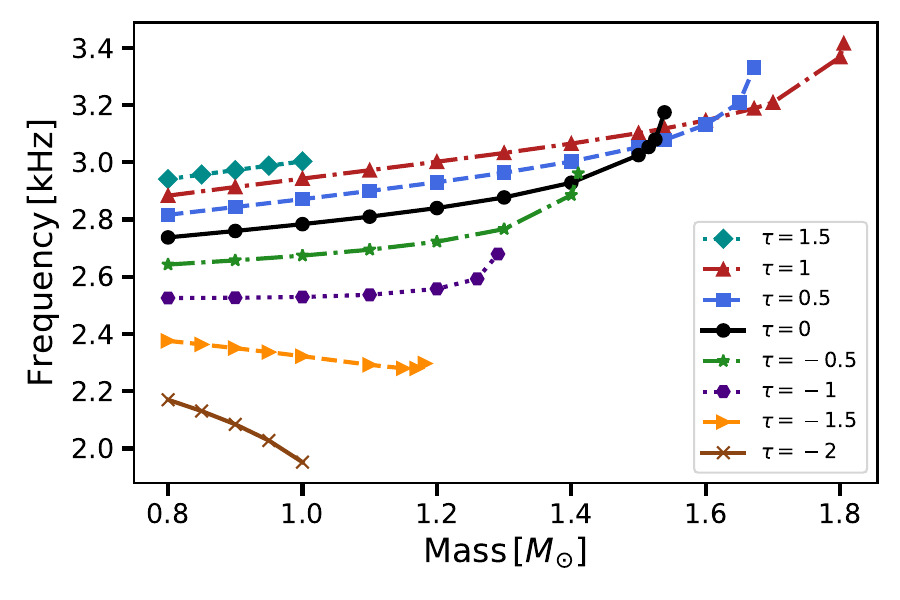}
	\caption{The variation of the $f$-mode frequency with the mass of quark stars for various values of the anisotropic strength for MIT bag EOS (left panel) and EOS-A (right panel).}
	\label{fmode-mass}
\end{figure*}

\begin{figure*}
	\centering
	\includegraphics[width=0.48\textwidth]{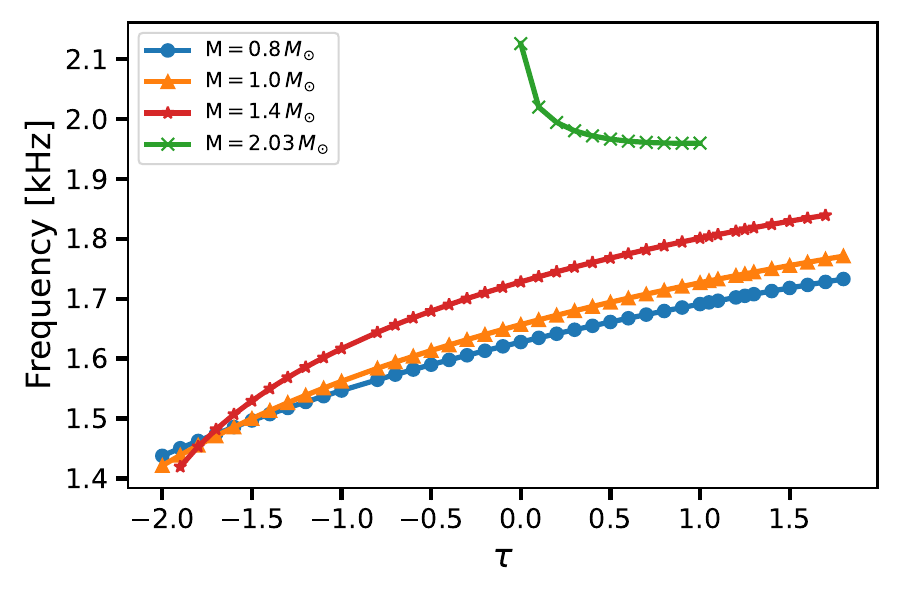}
	\hfill
	\includegraphics[width=0.48\textwidth]{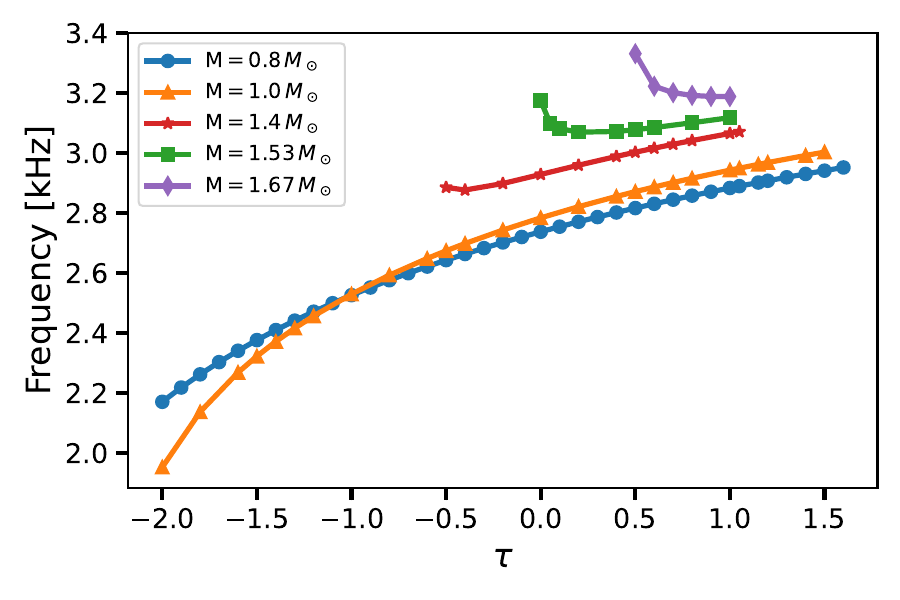}
	\caption{The variation of the $f$-mode frequency with the anisotropic strength for quark stars of fixed masses for MIT bag EOS (left panel) and EOS-A (right panel).}
	\label{f-tau}
\end{figure*}

Next, we investigate the effect of the anisotropy on the $f$-mode frequency. In \citetalias{mondal2024prd}, for different values of $\tau$, we found linear relations between the f-mode frequency and $\sqrt{\rho_{\rm avg}}$ where the slope and intercept of the lines depended on the chosen value of $\tau$. It is naturally intriguing to check whether any such general relation exists for quark stars. 

In Fig. \ref{freq_avden}, we plot the values of $\sqrt{\rho_{\rm avg}}$ along the abscissa and the corresponding values of the the f-mode frequency along the ordinate for various values of $\tau$ for the MIT bag EOS (left panel) and EOS-A (right panel). We see that for both EOSs, when $(\tau = 0)$, the frequency varies linearly with $\sqrt{\rho_{\rm avg}}$. For anisotropic stars, while the relationship remains nearly linear, the slope and intercept are noticeably influenced by the anisotropy. The effects are more pronounced for stars with $p_t < p_r$ $(\tau < 0)$ in both EOSs. This behavior can be mathematically expressed as:
\begin{equation}\label{MIT_freq}
	\mathcal{F}(\rho_\text{avg},\tau) \approx C(\tau) \sqrt{\rho_\text{avg}} + D(\tau), 
\end{equation}     
where $\mathcal{F}$ is the $f$-mode frequency, and $C(\tau)$ and $D(\tau)$ are functions of $\tau$ that capture the slope of the linear fit and the vertical axis intercept, respectively, in Fig.~\ref{freq_avden}. The explicit forms of $C(\tau)$ and $D(\tau)$, along with their dependence on $\tau$, are derived and discussed in detail in Sec.~\ref{sec:analytic}.  

In Fig. \ref{fmode-mass}, we plot the $f$-mode frequency as a function of the total mass of quark stars for different anisotropic strengths. The left panel corresponds to the MIT bag EOS, while the right panel represents results for EOS-A. We see that for $\tau \geq 0$ with the MIT bag EOS, the $f$-mode frequency exhibits a linear increase with mass for stars less massive than $2~{\rm M_\odot}$. However, for more massive stars, a nonlinear rise in frequency is observed. A similar linear increase in frequency is seen for $\tau > 0$ with EOS-A, but this behavior is limited to stars with relatively low masses of $1.4 \sim 1.5~{\rm M_\odot}$. Beyond this range, a nonlinear increase in frequency becomes prominent. For negative anisotropy $(\tau < 0)$, the trends are mostly similar to those observed for $\tau \geq 0$. With the MIT bag EOS, the linear increase in frequency persists only up to masses of approximately $1.5 \sim 1.6~{\rm M_\odot}$. In contrast, with EOS-A, this linear regime is restricted to lower masses, around $1.2 \sim 1.3~{\rm  M_\odot}$. Beyond these mass ranges, the frequency shows a nonlinear growth for $\tau = -0.5, -1.0$, and $-1.5$. For both EOSs, however, the behavior for $\tau = -2$ deviates significantly, the frequency decreases monotonically as the mass increases.

As mass is a fundamental and measurable parameter of stars, we next investigate the impact of anisotropy on the $f$-mode frequency for quark stars of fixed masses. For this purpose, we plot the $f$-mode frequency as a function of $\tau$ for several masses in Fig.~\ref{f-tau}, whose the left panel is for MIT bag EOS and the right panel is for EOS-A. For quark stars modeled with the MIT bag EOS, an increase in the anisotropic strength $\tau$ from $0$ to $1.7$ leads to a rise in the $f$-mode frequency, with the frequencies increasing by $6.16\%$ and $6.42\%$ for quark stars of $0.8~{\rm M_\odot}$ and $1.4~{\rm M_\odot}$, respectively. As $\tau$ decreases from $0$ to $-1.7$, the $f$-mode frequency decreases significantly, by $9.4\%$, $11.2\%$, and $14.2\%$ for quark stars with masses of $0.8~{\rm  M_\odot}$, $1~{\rm  M_\odot}$, and $1.4~{\rm  M_\odot}$, respectively. For a more massive quark star with a mass of $2.03~{\rm M_\odot}$, we observe a frequency decrease of $7.81\%$ as $\tau$ increases from $0$ to $1$. This behavior is attributed to the rapid decrease in $\sqrt{\rho_{\rm avg}}$ with increasing $\tau$, as shown in Fig.~\ref{avden-tau} and discussed earlier in this section. According to Eq.~(\ref{MIT_freq}), the $f$-mode frequency is directly proportional to $\sqrt{\rho_{\rm avg}}$. Thus, the observed rapid decrease in $\sqrt{\rho_{\rm avg}}$ for higher masses leads to a corresponding reduction in the frequency as $\tau$ increases. Notably, a similar trend has been reported for neutron stars in \citetalias{mondal2024prd}.

A similar trend is observed for quark stars modeled with EOS-A. For stars with masses of $0.8~{\rm M_\odot}$, $1~{\rm M_\odot}$, and $1.4~{\rm M_\odot}$, the $f$-mode frequency increases by $5.3\%$, $5.7\%$, and $4.66\%$, respectively, as $\tau$ increases from $0$ to $1$. Conversely, as $\tau$ decreases from $0$ to $-1$, the $f$-mode frequency decreases by $7.72\%$ for a $0.8~{\rm M_\odot}$ star and by $9.14\%$ for a $1~{\rm M_\odot}$ star. It is clearly visible from the right panel of Fig. \ref{f-tau}.

From right panel of Fig.~\ref{f-tau}, we also note that for a $1.53~ {\rm  M_\odot}$ quark star, the frequency initially drops by approximately $3.3\%$ as $\tau$ changes from $0$ to $0.2$, before increasing by $1.57\%$ as $\tau$ rises further from $0.2$ to $1$. This is a combined results of Fig. \ref{avden-tau} and Fig. \ref{freq_avden}. Fig. \ref{avden-tau} shows that for a fixed mass of the star, the value of  $\sqrt{\rho_{\rm avg}}$ decreases with the increase of $\tau$, and this decrease is more rapid for the lower values of $\tau$. The lines of $\sqrt{\rho_{\rm avg}} - \tau$ for different masses are not exactly same shapes. Similarly, from Fig. \ref{freq_avden}, we see that for a fixed value of $\tau$, $\mathcal{F}$ increases with the increase of $\sqrt{\rho_{\rm avg}}$ (except of $\tau=-2$) and for any given value of $\sqrt{\rho_{\rm avg}}$, $\mathcal{F}$ increases with the increase of $\tau$. The $\mathcal{F} - \sqrt{\rho_{\rm avg}}$ are not parallel as seen in \ref{freq_avden}. This complicated interplay between the frequency, average density, and the anisotropy manifest differently for stars of different masses as seen in Fig. \ref{f-tau}. This also motivated us to introduce an empirical relation between $\mathcal{F}$, $\sqrt{\rho_{\rm avg}}$, and $\tau$ as given in Eq. (\ref{MIT_freq}) as well as between $\mathcal{F}$, $M$, $R$, and $\tau$ as given in Eq. (\ref{eq:fitF_MRtau}). We will discuss Eq. (\ref{eq:fitF_MRtau}) in details afterward.

For more massive stars, such as a $1.67~{\rm M_\odot}$ quark star, the $f$-mode frequency decreases by $4.29\%$ as $\tau$ increases from $0.5$ to $1$. This behavior arises from the rapid decline in $\sqrt{\rho_{\rm avg}}$ with increasing $\tau$, as evident from the right panel of Fig. \ref{avden-tau}.

\subsection{Effect of anisotropy on damping time}

\begin{figure*}
	\centering
	\includegraphics[width=0.48\textwidth]{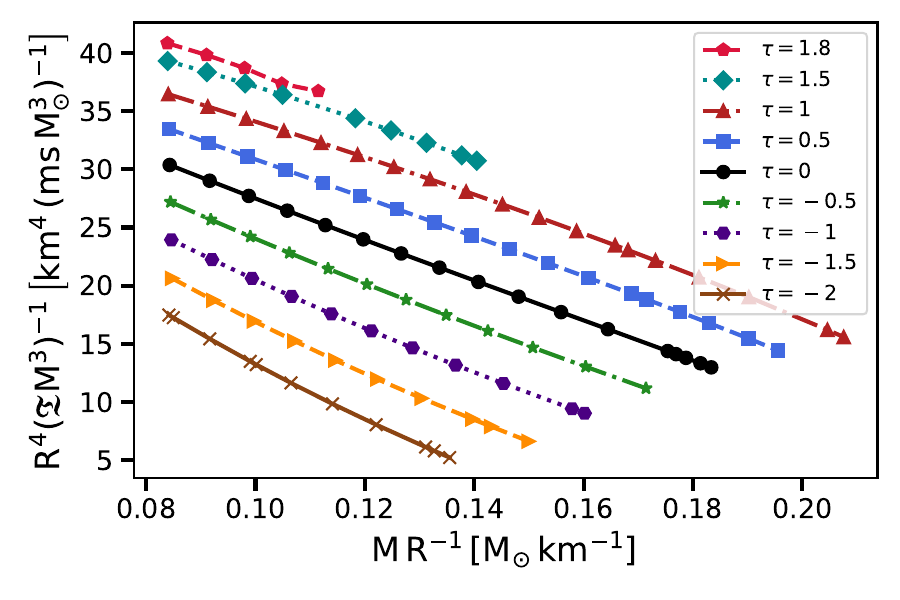}
	\hfill
	\includegraphics[width=0.48\textwidth]{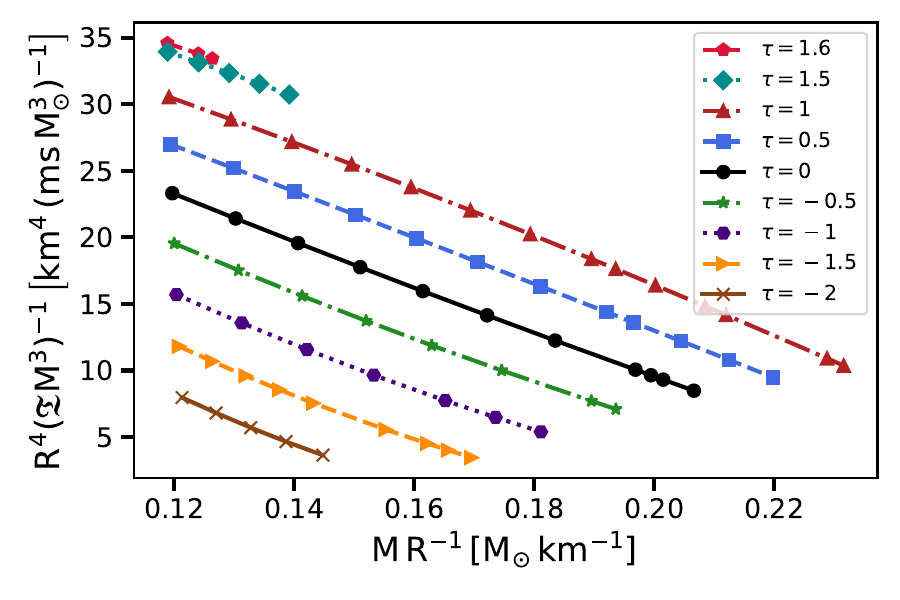}
	\caption{The variation of the inverse normalized damping time $(R^4 / \mathfrak{T} M^3)$ with the compactness $(M/R)$ for quark stars with different values of the anisotropic strength $\tau$. The left panel corresponds to the MIT bag EOS, while the right panel represents EOS-A.}
	\label{dt-R4M3}
\end{figure*}

We now examine the damping time of $f$-modes, which characterizes the dissipation of energy due to gravitational wave emission from the oscillations of compact stars. \citet{detweiler1975variational} demonstrated that the damping time $\mathfrak{T}$ of the $f$-mode scales proportionally with $R^4/M^3$, where $M$ and $R$ are the total mass and radius of the star, respectively. Later, \citet{andersson1998MNRAS} expressed the normalized inverse damping time, $( \mathfrak{T} M^3/R^4)^{-1}$, as a function of the compactness, $M/R$. Following this approach, we plot $( \mathfrak{T} M^3/R^4)^{-1}$ on the vertical axis against $M/R$ on the horizontal axis for both EOSs in Fig.~\ref{dt-R4M3}.

\begin{figure*}
	\centering
	\includegraphics[width=0.48\textwidth]{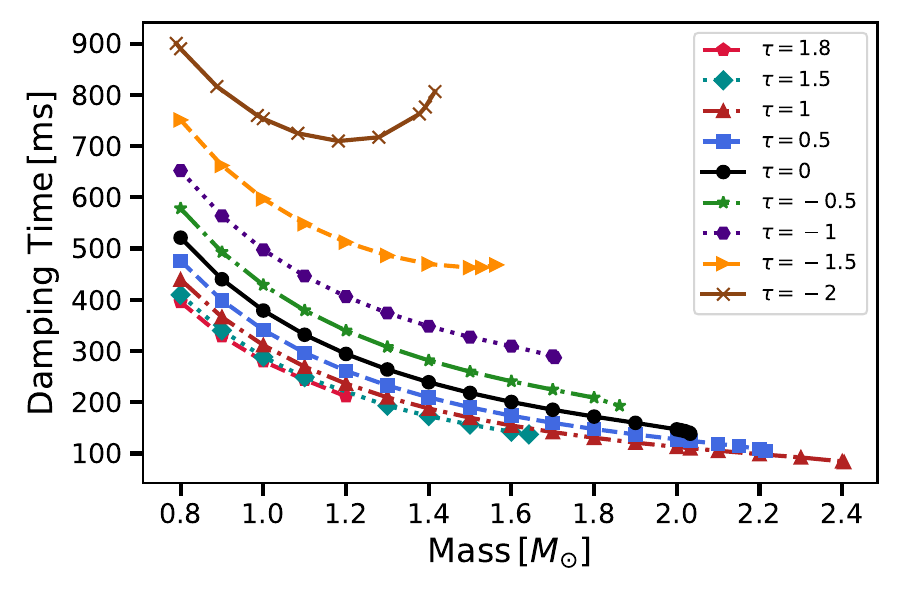}
	\hfill
	\includegraphics[width=0.48\textwidth]{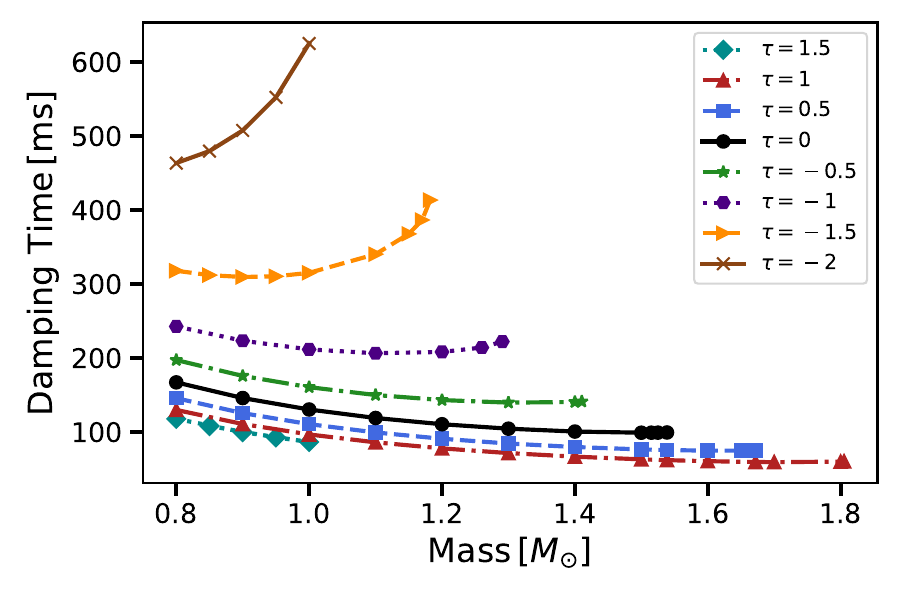}
	\caption{The variation of the damping time with mass of quark stars for various values of the anisotropic strength for MIT bag EOS (left panel) and EOS-A (right panel).}
	\label{dt-M}
\end{figure*}

\begin{figure*}
	\centering
	\includegraphics[width=0.48\textwidth]{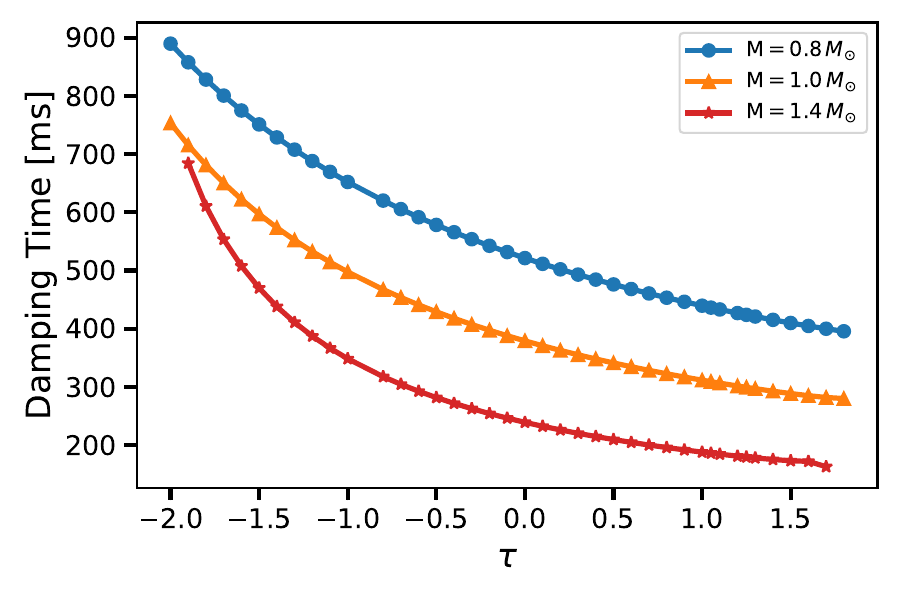}
	\hfill
	\includegraphics[width=0.48\textwidth]{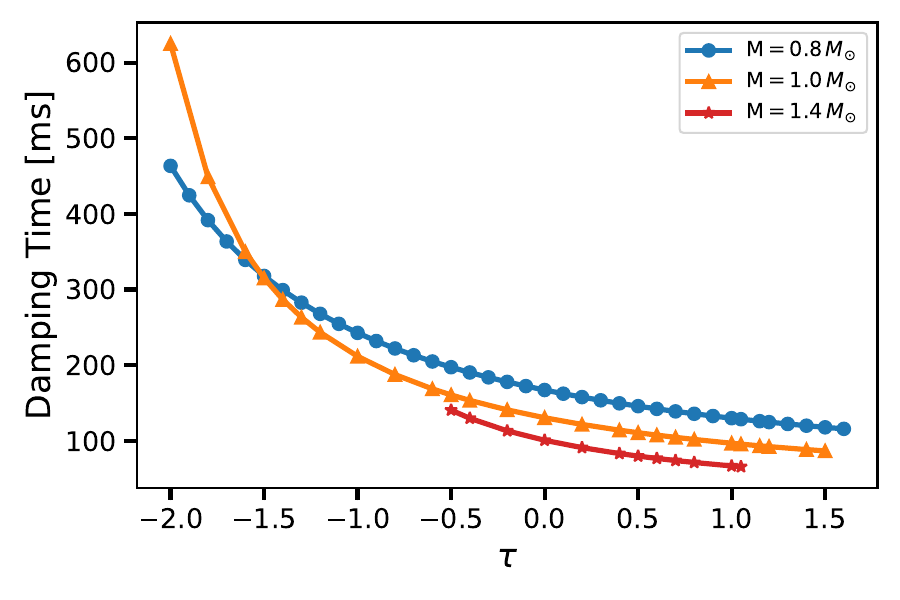}
	\caption{The variation of the damping time with the anisotropic strength for quark stars of fixed masses for MIT bag EOS (left panel) and EOS-A (right panel).}
	\label{dt-tau}
\end{figure*}

From Fig.~\ref{dt-R4M3}, it is evident that the normalized inverse damping time decreases linearly with the increase of $M/R$. The effect of the anisotropy is embedded in the slope and intercept of the fit. Consequently, we approximate the normalized damping time analytically using the following expression

\begin{equation}\label{analytic:damping}
	\frac{R^4}{\mathfrak{T} M^3} \approx J(\tau)\frac{M}{R} + K(\tau),
\end{equation}  where $J(\tau)$ correspond to the slope and $K(\tau)$ corresponds to the intercept on the vertical axis of the fitted $(\mathfrak{T} M^3/R^4)^{-1}$ versus $(M/R)$ lines. A more detailed discussion of this analytical function is provided in Sec.~\ref{sec:analytic}.

To further understand the impact of the anisotropy, we examine how the damping time varies with the mass of quark stars. In Fig.~\ref{dt-M}, the damping time is plotted on the vertical axis, while the total mass is shown on the horizontal axis for both of the EOSs. For $\tau \geq 0$, the damping time decreases monotonically as the mass of the quark star increases for both EOSs. For the MIT bag EOS, a steep increase in damping time is observed for massive quark stars when $\tau \leq -1.5$ as seen in the left panel of Fig.~\ref{dt-M}. The right panel of Fig.~\ref{dt-M} shows similar behavior for EOS-A when $\tau \leq -1$. Intriguingly, for $\tau = -2$, the damping time increases monotonically with the mass of quark stars for the EOS-A, while no such monotonic trend is seen in the MIT bag EOS.

To investigate the influence of anisotropy on the damping time of quark stars with specific masses, we plot the damping time as a function of anisotropic strength $\tau$ for stars with masses of $0.8~{\rm M_\odot}$, $1~{\rm M_\odot}$, and $1.4~{\rm M_\odot}$ for both of the EOSs in Fig.~\ref{dt-tau}. For the MIT bag EOS (the left panel of Fig.~\ref{dt-tau}), we observe a decrease in damping time as $\tau$ increases from $0$ to $1.7$, amounting to $23.29\%$, $25.65\%$, and $31.90\%$ for quark stars with masses of $0.8~{\rm M_\odot}$, $1~{\rm M_\odot}$, and $1.4~{\rm M_\odot}$, respectively. Conversely, when $\tau$ decreases from $0$ to $-1.7$, the damping time increases significantly, by $53.56\%$, $71.55\%$, and $131.48\%$ for the same respective masses.

Similarly, for EOS-A (the right panel of Fig.~\ref{dt-tau}), we observe a $29.48\%$ and $33.59\%$ reduction in damping time for quark stars with masses of $0.8~{\rm M_\odot}$ and $1~{\rm M_\odot}$, respectively, as $\tau$ increases from $0$ to $1.5$. On the other hand, as $\tau$ decreases from $0$ to $-1.5$, the damping time increases significantly by $90.15\%$ and $141.40\%$ for the same respective masses. For a quark star with a mass of $1.4~{\rm M_\odot}$, the damping time decreases by $34.69\%$ as $\tau$ increases from $0$ to $1.05$. For $\tau < 0$, we do not get stable, physical quark stars as massive as $1.4~{\rm M_\odot}$ (see right panels of Figs. \ref{M-Rhoc} and \ref{M-R}).

The preceding discussions make it clear that for both EOSs, the frequency and the damping time of the $f$-mode oscillations are significantly affected by the values of the mass, the radius, and the anisotropic strength of quark stars. While the dependence of both $\mathcal{F}$ and $\mathfrak{T}$ on these parameters follow similar overall trends, the specific ranges of values of the parameters differ between the two EOSs. For the MIT bag EOS, the $f$-mode frequency varies from $1.3~\text{kHz}$ to $2.3~\text{kHz}$, depending on the mass, radius, and anisotropy of the star, whereas for EOS-A, it spans a slightly higher range of $1.8~\text{kHz}$ to $3.4~\text{kHz}$. A similar distinction is observed for the damping time, which lies between $83~\text{ms}$ and $900~\text{ms}$ for the MIT bag EOS, while for EOS-A, it ranges from $60~\text{ms}$ to $761~\text{ms}$. These results highlight how the equation of state, along with the structural properties of the star and the degree of the anisotropy, governs the $f$-mode oscillation characteristics of quark stars.

\section{Expressions for the frequency and the damping time as a function of mass, radius, and the anisotropic strength}\label{sec:analytic}

In this section, we derive semi-empirical expressions for the frequency and the damping time of the $f$-mode oscillations as functions of the intrinsic properties of quark stars, such as total mass, radius, and anisotropic strength.

\subsection{Expression for the frequency}
\label{subsec:fitexpF}

We have already seen in Fig. \ref{freq_avden}, that the frequency of the f-mode oscillation is almost linearly  proportional to the square root of the average density of the quark star, and the values of $\tau$ affect the slopes and intercepts of the $\mathcal{F} - \sqrt{\rho_{\rm avg}}$ lines. The approximate functional form was given in Eq. (\ref{MIT_freq}), which can be re-written as:

\begin{equation}
\label{eq:fitF_MRtau}
	\mathcal{F}(M, R, \tau) \approx C(\tau) \sqrt{\frac{3}{4 \pi}}\sqrt{\frac{M}{R^3}} + D(\tau),
\end{equation} where the definition $\rho_{\rm avg} = M [(4/3) \pi R^3]^{-1}$ has been used.

To obtain expressions for $C(\tau)$ and $D(\tau)$, we first compute the values of $\mathcal{F}$ for a range of stable quark star masses, from $0.8~{\rm M_\odot}$ to the maximum stable mass, for various fixed values of $\tau$. We used $\tau$ in the range of $-2$ to $1.8$ for MIT bag EOS and in the range of $-2$ to $1.6$ for EOS-A. For each value of $\tau$, we performed a linear fit between $\mathcal{F}$ and $\sqrt{3M/(4 \pi R^3)}$, the slope of each fit giving the value of $C$ for that $\tau$ while the intercept of the fit giving the value of $D$. In this step, we have used the values of $M$ in the unit of ${\rm M_{\odot}}$, the values of $R$ in the unit of kilometers (km), and the values of $\mathcal{F}$ in the unit of kilohertz (kHz).

Afterwards, for the two EOSs separately, we collect various values of $C$, $D$, and corresponding $\tau$ and perform polynomial fits (up to the third order) as:
\begin{align}
	\label{eq:fitC}
	C(\tau) = c_0 + c_1 \tau + c_2 \tau^2 + c_3 \tau^3,\\
		\label{eq:fitD}
	D(\tau) = d_0 + d_1 \tau + d_2 \tau^2 + d_3 \tau^3,
\end{align}  
where $c_0$, $c_1$, $c_2$, and $c_3$ are the fitting parameters for $C(\tau)$, and $d_0$, $d_1$, $d_2$, and $d_3$ are the fitting parameters for $D(\tau)$. The best fit parameters for the two EOSs are given in Tables \ref{fitparam1} and \ref{fitparam2}.

\begin{table}
	\centering
	\caption{Fitting parameters for $C(\tau)$ for two EOSs. The parameters $c_0$, $c_1$, $c_2$, and $c_3$ correspond to the coefficients obtained from the fit.}
	\label{fitparam1}
	\begin{tabular}{lcccc}
		\hline\hline
		EOS     & $c_0$       & $c_1$       & $c_2$         & $c_3$       \\ \hline
		MIT bag & 121.1240     & 18.4186    & $-2.1998$    & 13.1095    \\ 
		EOS-A   & 87.8584    & 12.3431   & $-6.7491$    & 28.0644    \\ 
		\hline\hline
	\end{tabular}
\end{table}

\begin{table}
	\centering
	\caption{Fitting parameters for $D(\tau)$ for two EOSs. The parameters $d_0$, $d_1$, $d_2$, and $d_3$ correspond to the coefficients obtained from the fit.}
	\label{fitparam2}
	\begin{tabular}{lcccc}
		\hline\hline
		EOS     & $d_0$       & $d_1$       & $d_2$         & $d_3$       \\ \hline
		MIT bag & $-0.1817 $  & $-0.1777$    & 0.0261     & $-0.1995 $ \\ 
		EOS-A   & 0.5130      & $-0.0936$   & 0.1472   & $-0.7172$    \\ 
		\hline\hline
	\end{tabular}
\end{table}

We have also checked the quality of the fit by calculating the values of the coefficient of the determination $\mathcal{R}^2$, which is defined as \cite{kvalseth1985tas, godfrey1964nrlsq}:
\begin{equation}\label{R2}
	\mathcal{R}^2 = 1 - \frac{\sum\limits_{i}  (Y_n - Y_f)^2}{\sum\limits_{i}(Y_n - \bar{Y})^2},
\end{equation}
where $Y_n$ is the numerically calculated value of the quantity $Y$ (here the $f$-mode frequency), $Y_f$ is the corresponding value obtained from the fit, and $\bar{Y}$ is the mean of the numerically calculated values. An $\mathcal{R}^2$ value greater than $0.95$ is generally considered excellent and indicates that the fit reliably captures the functional dependence. For the MIT bag EOS, the $\mathcal{R}^2$ value is $0.9971$, while for EOS-A, it is $0.9972$, indicating an excellent fit in both cases.

\begin{figure*}
	\centering
	\includegraphics[width=0.48\textwidth]{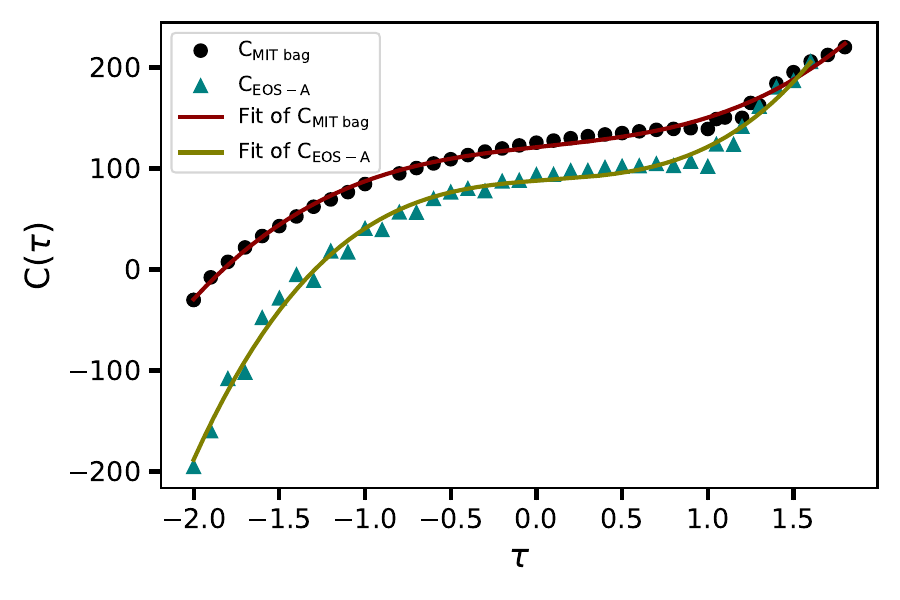}
	\hfill
	\includegraphics[width=0.48\textwidth]{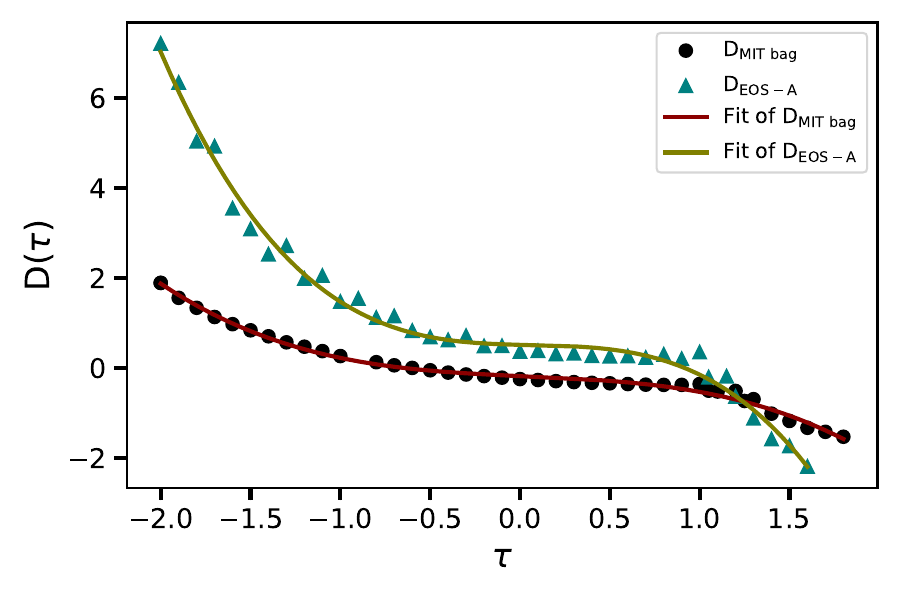}
	    \caption{The variation of $C(\tau)$ (left panel) and $D(\tau)$ (right panel) with anisotropic strength $\tau$ for quark stars modeled using the MIT bag EOS and EOS-A. The discrete points represent the numerically obtained values for the slope $C$ and intercept $D$ of the linear fit in the $\mathcal{F} - \sqrt{\rho_{\rm avg}}$ relation, while the solid lines denote the best-fit polynomial approximations. The subscripts in $C$ and $D$ specify the corresponding EOS.}
	\label{CD-fit}
\end{figure*}

For visual representations, in Fig. \ref{CD-fit} we plot the values $\tau$ along the abscissa and the values of $C$ (in the left panel) and $D$ (in the right panel) along the ordinate for both EOSs. We use discrete points for the values obtained from fits of $\mathcal{F}$ and $\sqrt{3M/(4 \pi R^3)}$ and use lines when we use Eqs. (\ref{eq:fitC}) and (\ref{eq:fitD}) with the values of the coefficients as given in Tables \ref{fitparam1} and \ref{fitparam2}.

From the plots, we observe that the slope $C(\tau)$ is a monotonically increasing function of $\tau$, while the intercept $D(\tau)$ is a monotonically decreasing function. Another noteworthy feature is that the effects of anisotropy become more pronounced for $|\tau| \gtrsim 1$.

\subsection{Expression for the damping time}
\label{subsec:fitexpT}

We have already seen in Fig. \ref{dt-R4M3}, the normalized inverse damping time $(\mathfrak{T} M^3/R^4)^{-1}$, decreases linearly with the increase of $M/R$, and the values of $\tau$ affects the slope and intercepts of the fitted lines. The approximate functional form was given in Eq. (\ref{analytic:damping}). Here we aim to obtain expressions for $J(\tau)$ and $K(\tau)$. 

First, we compute the values of $(\mathfrak{T} M^3/R^4)^{-1}$ for a range of stable quark star masses, the range of the mass and the anisotropy parameter being the same as in subsection \ref{subsec:fitexpF}. Then, for each value of $\tau$, we performed a linear fit between $(\mathfrak{T} M^3/R^4)^{-1}$ and the compactness $M/R$. The slope of each fit giving the value of $J$ for that $\tau$ while the intercept of the fit giving the value of $K$. Similar to the earlier subsection, in this step, we have used the values of $M$ in the unit of ${\rm M_{\odot}}$, the values of $R$ in the unit of kilometers (km), and the values of $\mathfrak{T}$ in the unit of milliseconds (ms).

At the next step, for the two EOSs separately, we collect various values of $J$, $K$, and corresponding $\tau$ and perform polynomial fits (up to the fourth order) as:
\begin{align}
\label{eq:fitJ}
	J(\tau) = j_0 + j_1 \tau + j_2 \tau^2 + j_3 \tau^3 + j_4 \tau^4, \\
	\label{eq:fitK}
	K(\tau) = k_0 + k_1 \tau + k_2 \tau^2 + k_3 \tau^3 + k_4 \tau^4 ,
\end{align}

where $j_0$, $j_1$, $j_2$, $j_3$, and $j_4$ are the fitting parameters for $J(\tau)$, while $k_0$, $k_1$, $k_2$, $k_3$, and $k_4$ are the fitting parameters for $K(\tau)$. The best fit parameters for the two EOSs are given in Tables \ref{fitparam3} and \ref{fitparam4}. The accuracy of these fits is evaluated using the coefficient of determination, $\mathcal{R}^2$, defined in Eq. (\ref{R2}). We obtain $\mathcal{R}^2 = 0.9988$ for the MIT bag EOS and $\mathcal{R}^2 = 0.9869$ for EOS-A, indicating an excellent agreement between the fitted expressions and the numerical results.

\begin{table*}
	\centering
	\caption{Fitting parameters for $J(\tau)$ for two EOSs. The parameters $j_0$, $j_1$, $j_2$, $j_3$, and $j_4$ correspond to the coefficients obtained from the fit.}
	\label{fitparam3}
	\begin{tabular}{lccccc}
		\hline\hline
		EOS     & $j_0$       & $j_1$       & $j_2$         & $j_3$ & $j_4$      \\ \hline
		MIT bag & $-175.9430$     & 15.0989    & $-3.5854$    & 1.4289  & $-0.3524$ \\ 
		EOS-A   & $-171.4760$    & $-8.1686$    & $-2.6170$    & 5.9472   & 2.1901 \\ 
		\hline\hline
	\end{tabular}
\end{table*}

\begin{table*}
	\centering
	\caption{Fitting parameters for $K(\tau)$ for two EOSs. The parameters $k_0$, $k_1$, $k_2$, $k_3$, and $k_4$ correspond to the coefficients obtained from the fit.}
	\label{fitparam4}
	\begin{tabular}{lccccc}
		\hline\hline
		EOS     & $k_0$       & $k_1$       & $k_2$         & $k_3$ & $k_4$      \\ \hline
		MIT bag & 45.1012      & 5.4740     & 0.2597       & $-0.2780$     & $-0.0175$ \\ 
		EOS-A   & 43.7257      & 8.6682     & 0.2338       & $-0.8676$     & $-0.3018$ \\ 
		\hline\hline
	\end{tabular}
\end{table*}

\begin{figure*}
	\centering
	\includegraphics[width=0.48\textwidth]{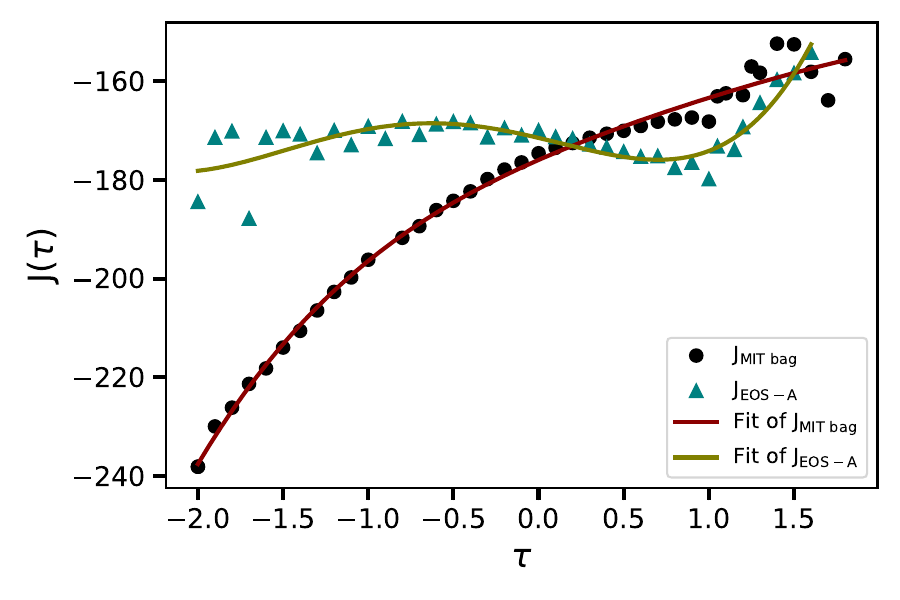}
	\hfill
	\includegraphics[width=0.48\textwidth]{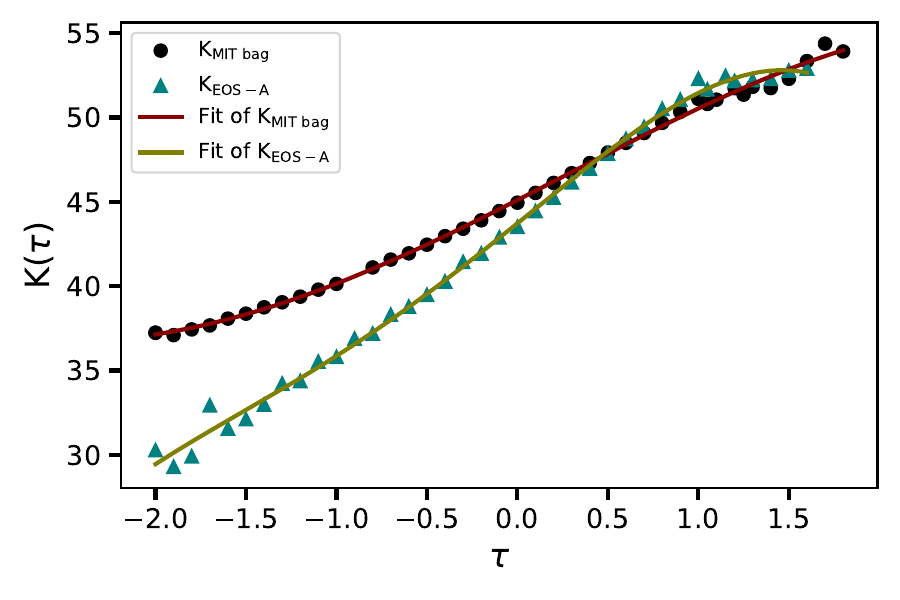}
	\caption{The variation of $J(\tau)$ (left panel) and $K(\tau)$ (right panel) with anisotropic strength $\tau$ for quark stars modeled using the MIT bag EOS and EOS-A. The discrete points represent the numerically obtained values for the slope $J$ and intercept $K$ of the linear fit in the $(\mathfrak{T} M^3/R^4)^{-1}$ versus $M/R$ relation, while the solid lines denote the best-fit polynomial approximations. The subscripts in $J$ and $K$ specify the corresponding EOS.}
	\label{JK-fit}
\end{figure*}

For visual representations, in Fig. \ref{JK-fit}, we plot the values $\tau$ along the abscissa and the values of $J$ (in the left panel) and $K$ (in the right panel) along the ordinate for both EOSs. We use discrete points for the values obtained from fits of $(\mathfrak{T} M^3/R^4)^{-1}$ and $M/R$ and use lines when we use Eqs. (\ref{eq:fitJ}) and (\ref{eq:fitK}) with the values of the coefficients as given in Tables \ref{fitparam3} and \ref{fitparam4}.

\section{Summary and Conclusion}\label{conclusion}

In this study, we have investigated some properties of anisotropic quark stars in full general relativity, considering two distinct equations of state (EOSs) to model quark matter, the MIT bag model, representing non-interacting quark matter, and EOS-A, which incorporates inter-quark interactions through a modified Richardson potential. We have studied structures of stable quark stars and associated parameters like the total mass and the radius, as well as the $f$-mode oscillations and associated properties like the frequency and the damping time of the oscillations.

Our analysis confirms that the $f$-mode frequency is linearly proportional to the square root of the average density of quark stars, with the explicit effect of the anisotropy imprinted on the slope and the intercept of the $\mathcal{F}-\sqrt{\rho_{\rm avg}}$ relation. A similar trend has been previously observed for neutron stars, as shown in \citetalias{mondal2024prd}. We also found that for $\tau \geq 0$, where the tangential pressure exceeds the radial pressure, the $f$-mode frequency increases monotonically with the mass, and a rapid growth observed in more massive stars. A similar trend is seen in general for $\tau < 0$, except for $\tau \sim -2$, where the frequency decreases monotonically with the increase of the  mass. Overall, we find that the $f$-mode frequency of quark stars varies within the range $1.3~\text{kHz}$ to $3.5~\text{kHz}$, depending on the EOS and the degree of the anisotropy present in the star.

Additionally, we observed that the normalized damping time is proportional to the compactness of the star, with the effects of the anisotropy manifesting in both the slope and the intercept of the corresponding fits. Furthermore, for a fixed mass, the damping time decreases as the tangential pressure increases relative to the radial pressure, while it significantly increases for configurations where the tangential pressure decreases with respect to the radial pressure. The magnitude of these changes varies with the mass and is described in detail in Sec.~\ref{result:fmode}.

We have also derived semi-empirical expressions for the frequencies and the damping times of the $f$-mode oscillations as functions of the mass, the radius, and the anisotropic strength. We find that for both EOSs, the dependence of frequency on $\tau$ is well approximated by a third-degree polynomial of $\tau$, while the damping time is fitted by a fourth-degree polynomial in $\tau$. These relations and the fit parameters can be used for other values of $\tau$ within the chosen range for the stable quark stars obtained with the EOSs used in this work.

In future work, this study can be extended by exploring additional EOSs and tuning the parameters of the equations of state considered here. This approach would enable the formulation of empirical relations for the frequency and the damping time of the $f$-mode oscillations of the anisotropic quark stars. If $f$-mode oscillations of compact stars are detected observationally, such relations could provide valuable constraints on the EOS of the dense matter and the degree of anisotropy within quark stars.

\section{Data Availability}\label{data}

No observational data have been used in this work. Numerical results and codes underlying this article will be shared on reasonable request to the corresponding author.

\section{Acknowledgements }\label{ack}

The authors thank the anonymous referee for constructive suggestions on the first version of the manuscript.

\bibliography{main}
\end{document}